\pdfoutput=1
\documentclass[runningheads]{llncs}
\usepackage{graphicx}
\usepackage{amsmath}
\usepackage{amssymb}
\usepackage{marvosym}

\begin{document}
\title{Transformer Network for Significant Stenosis Detection in CCTA of Coronary Arteries}
\titlerunning{Significant Stenosis Detection via Transformer Network}
\author{Xinghua Ma \and Gongning Luo\textsuperscript{(\Letter)} \and Wei Wang\textsuperscript{(\Letter)} \and Kuanquan Wang\textsuperscript{(\Letter)}}
\authorrunning{X. Ma et al.}
\institute{Harbin Institute of Technology, Harbin, China \\
\email{\{luogongning,wangwei2019,wangkq\}@hit.edu.cn}}
\maketitle
\begin{abstract}
Coronary artery disease (CAD) has posed a leading threat to the lives of cardiovascular disease patients worldwide for a long time. Therefore, automated diagnosis of CAD has indispensable significance in clinical medicine. However, the complexity of coronary artery plaques that cause CAD makes the automatic detection of coronary artery stenosis in Coronary CT angiography (CCTA) a difficult task. In this paper, we propose a Transformer network (TR-Net) for the automatic detection of significant stenosis (i.e. luminal narrowing $> 50\%$) while practically completing the computer-assisted diagnosis of CAD. The proposed TR-Net introduces a novel Transformer, and tightly combines convolutional layers and Transformer encoders, allowing their advantages to be demonstrated in the task. By analyzing semantic information sequences, TR-Net can fully understand the relationship between image information in each position of a multiplanar reformatted (MPR) image, and accurately detect significant stenosis based on both local and global information. We evaluate our TR-Net on a dataset of 76 patients from different patients annotated by experienced radiologists. Experimental results illustrate that our TR-Net has achieved better results in ACC (0.92), Spec (0.96), PPV (0.84), F1 (0.79) and MCC (0.74) indicators compared with the state-of-the-art methods. The source code is publicly available from the link (\url{https://github.com/XinghuaMa/TR-Net}).
	
\keywords{Coronary artery stenosis  \and Transformer \and Coronary CT angiograph \and Automatic detection}
\end{abstract}
\section{Introduction}
Coronary artery disease (CAD), as a common cardiovascular disease, has been a leading threat to human health around the world\cite{luo2020dynamically,luo2020commensal}. It is caused by atherosclerotic plaques in the main blood supply branches of coronary artery trees and induces the stenosis or blockage of blood vessels, resulting in the symptoms of heart disease, such as myocardial ischemia, angina pectoris and heart failure\cite{mendis2015organizational}. Coronary CT angiography (CCTA) is a practical non-invasive vascular imaging technique, which plays an important role in the perioperative period of interventional treatment of CAD. Analyzing grades of stenosis accurately through the pathological information in CCTA scans is essential for clinical applications related to CAD\cite{dewey2007coronary}.

\begin{figure}
	\includegraphics[width=\textwidth]{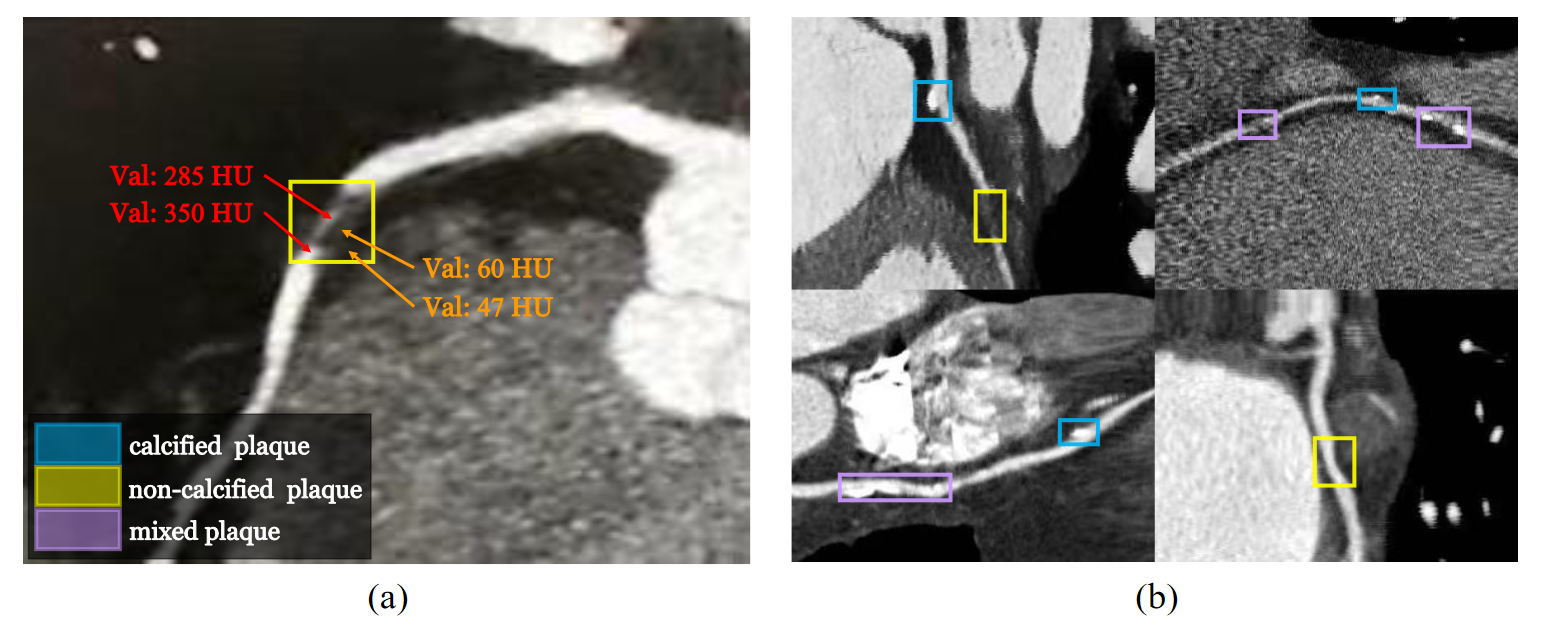}
	\caption{The HU values of non-calcified plaques are similar to adjacent tissues (a), and types of plaques are complicated and shapes vary from different plaques (b).} \label{fig_CAD}
\end{figure}

In recent years, computer vision technology has been used to detect coronary artery stenosis through CCTA scans of patients to assist clinicians in the diagnosis of CAD. However, it is challenging to realize the automatic diagnosis of CAD, because the contrast of the HU values among adjacent tissues and structures is low. Besides, types of plaques that cause coronary artery stenosis are complicated and there is no shape feature that can be used to strictly describe plaques, as shown in Fig. \ref{fig_CAD}.

Originally, most of the proposed methods were semi-automatic detection methods that require a lot of manual interaction\cite{kiricsli2013standardized}. And then, several machine learning-based methods have been proposed to describe changes in coronary lumens to quantify stenosis\cite{sankaran2016hale}. To a certain extent, these methods demonstrated that the geometric information of coronary lumen is considerable in clinical medicine. Several deep learning-based methods have been reported for the automatic detection of coronary artery stenosis in recent related literature\cite{shen2017deep,shin2016deep}. These works mainly employed the model combining convolutional neural network (CNN) and recurrent neural network (RNN) to complete the task. Zreik et al. achieved the detection of coronary artery plaque and stenosis by a recurrent convolutional neural network (RCNN). Particularly, they firstly reconstructed multiplanar reformatted (MPR) images based on the centerlines of coronary arteries. Next, they employed a 3D-CNN to extract features from small volumes and achieved the classification of two tasks using an RNN\cite{zreik2018recurrent}. Denzinger et al. improved the network structure of RCNN and predicted significant stenosis (i.e. luminal narrowing $> 50\%$) with the combination of deep learning approach and radiomic features\cite{denzinger2019coronary}. Also, Tejero-de-Pablos et al. extracted features from five views of coronary arteries and employed a Fisher vector to predict the classification probability of significant stenosis according to the features of varied views\cite{tejero2019texture}.

Although RNN can capture the dependencies between semantic features in a single direction to a certain extent, the global intervention of coronary artery branches to detect coronary artery stenosis is hardly considered in related work. To ensure that the model can learn the semantic features of entire coronary artery branches before local coronary artery stenoses are detected, we introduce Transformer into our method to analyze MPR images from a different perspective from others. Transformer is a type of deep neural network based on self-attention module\cite{vaswani2017attention}, which was invented to solve related tasks in the natural language processing (NLP) field\cite{han2020survey}. Transformer employs an attention mechanism to capture global context information to establish a long-distance dependence on the target, thereby extracting more ponderable features. In recent years, researchers in the computer vision field have continuously tapped its application potential in computer vision tasks\cite{carion2020end,wang2018non}.

\begin{figure}
	\includegraphics[width=\textwidth]{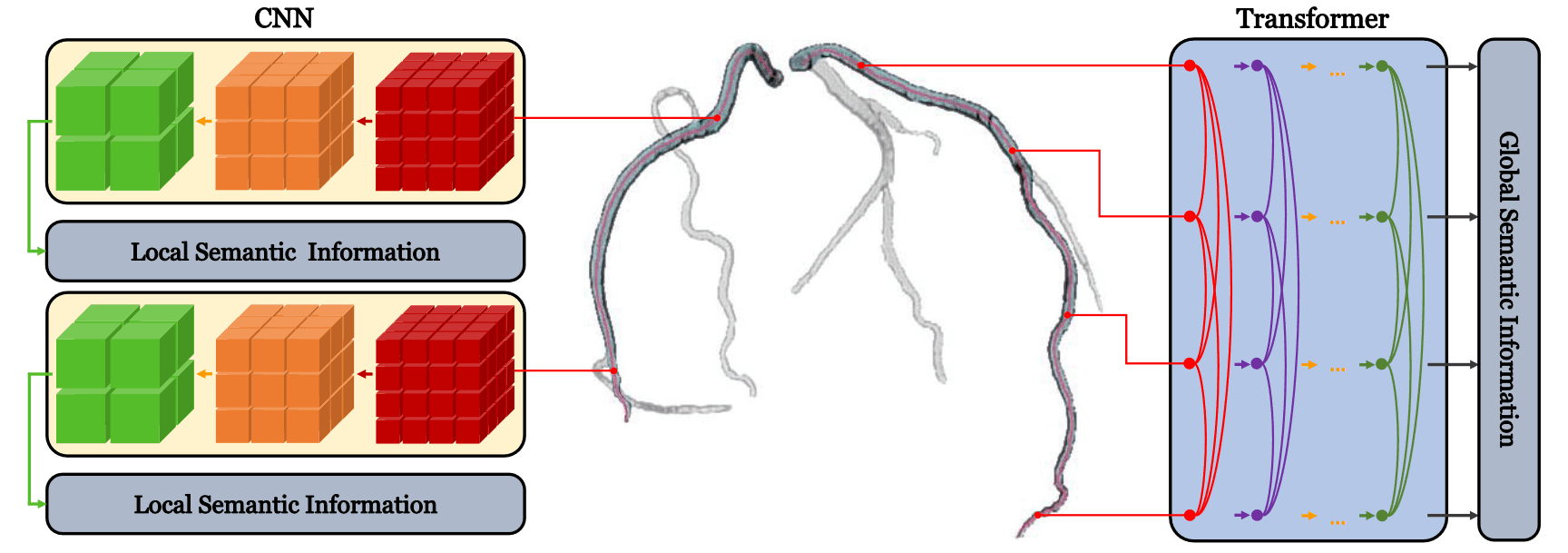}
	\caption{CNN for obtaining local semantic information and Transformer for obtaining global semantic information.} \label{fig_CNN_Transformer}
\end{figure}

In this work, we propose a novel Transformer Network (TR-Net) to detect significant stenosis in MPR images. The proposed TR-Net combines CNN and Transformer. As shown in Fig. \ref{fig_CNN_Transformer}, the former has a relatively large advantage in extracting local semantic information, while the latter can more naturally associate global semantic information. We employ a shallow 3D-CNN to extract local semantic features of coronary arteries. The shallow CNN enables the model to obtain the semantic information of each position in an MPR image while ensuring the efficiency of our model. Then, Transformer encoders are used to analyze feature sequences, which can mine the underlying dependence of local stenosis on each position of a coronary artery.

Our main contributions can be summarized as follows: (1) To achieve a more accurate diagnosis of coronary artery stenosis, we introduce Transformer to solve the challenging problem. To the best of our knowledge, this is the first attempt employing Transformer structure to complete the task of detecting coronary artery stenosis. (2) The proposed TR-Net can effectively integrate local and global information to detect significant stenosis. Experimental results illustrate that the proposed method has higher accuracy than state-of-the-art methods.

\section{Method}
\begin{figure}
	\includegraphics[width=\textwidth]{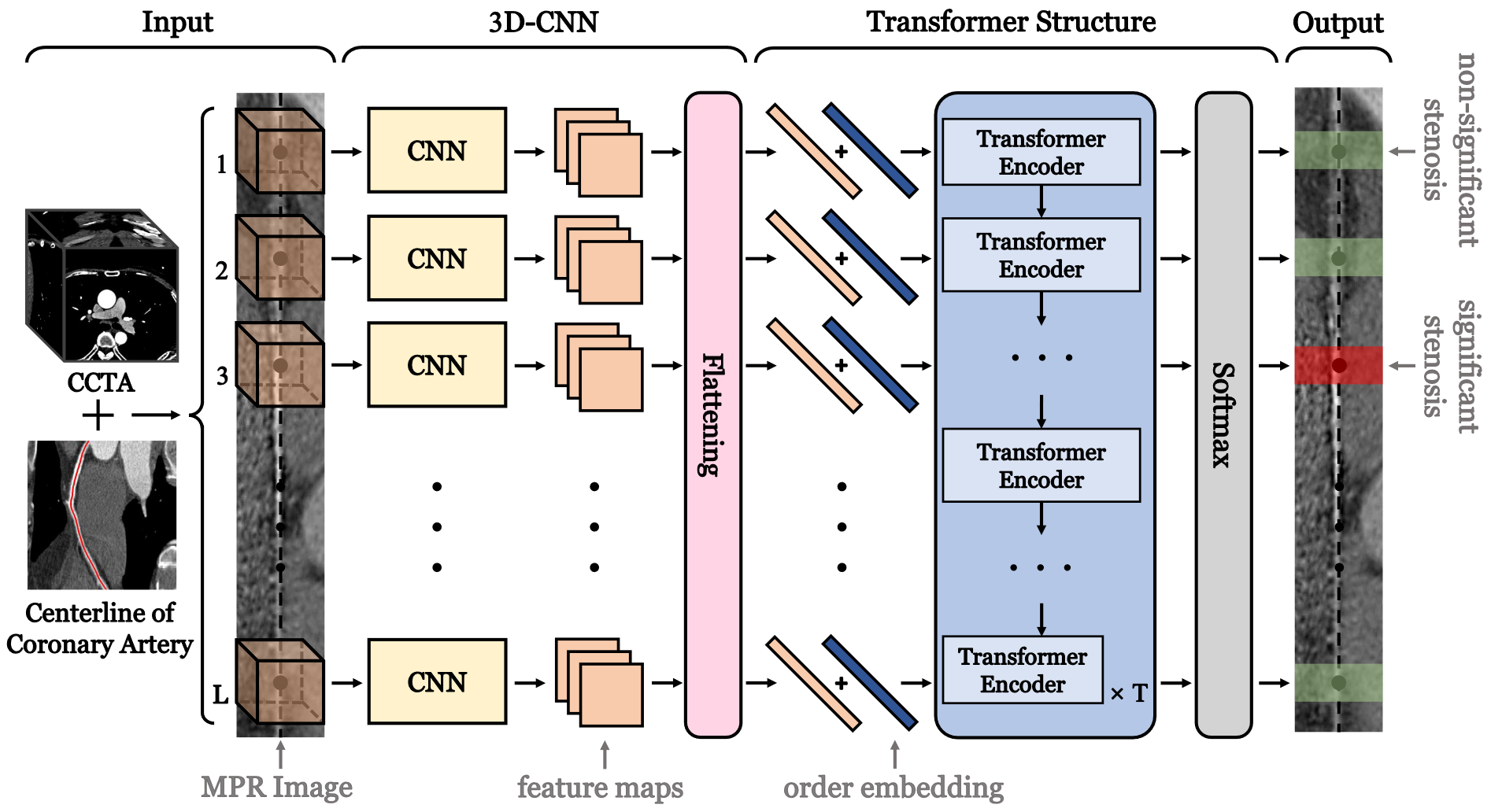}
	\caption{Transformer network (TR-Net).} \label{fig_TR-Net}
\end{figure}

In this section, we detail the proposed TR-Net for significant stenosis detection. Fig. \ref{fig_TR-Net} illustrates the architecture of TR-Net. TR-Net mainly consists of two components. One part is the 3D-CNN used to extract local semantic features at different positions of a coronary artery. The other part is Transformer structure used to associate the local feature maps of each position, analyzing the dependence of different positions, and classifying the significant stenosis at each position, as shown in Fig. \ref{fig_CNN_Transformer}.

\subsection{Semantic feature extraction for local cubic volumes}
For a certain locality of a coronary artery, the detail of the partial image is an indispensable reference basis for doctors when diagnosing CAD. This is also a prerequisite for our method to detect coronary artery stenosis. To efficiently extract local semantic features of coronary arteries, we design a shallow 3D-CNN as the first part of our method. The shallow CNN can not only prevent the overfitting of semantic information but also improve the efficiency of the model.

The input of our method is a coronary artery MPR image. We employ the voxels on the centerline of the coronary artery as center points to select cubic volumes from the MPR image, and the side length of cubic volumes is $N$ voxels. Then, we arrange these cubic volumes into a sequence of length $L$ according to the topological relationship of the coronary artery centerline. The semantic features of cubic volumes extracted by 3D-CNN are treated as the input of Transformer structure.

\begin{figure}
	\includegraphics[width=\textwidth]{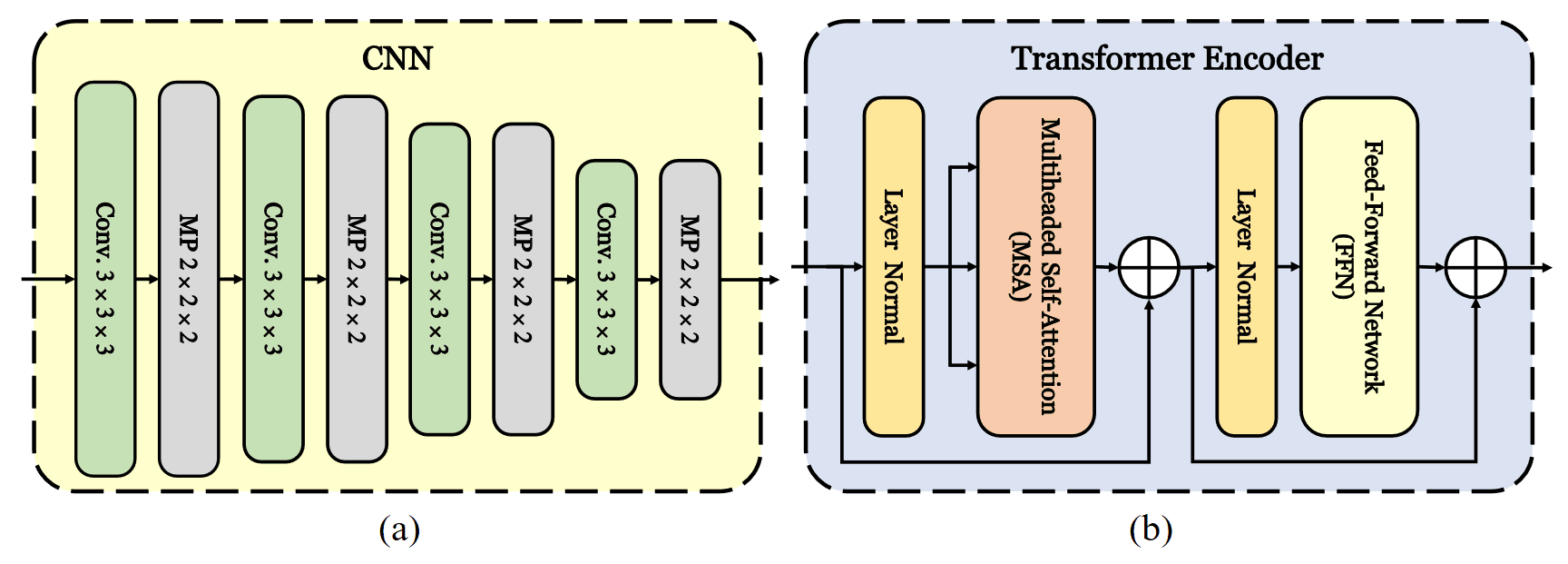}
	\caption{(a) The shallow 3D-CNN. (b) Transformer encoder.} \label{fig_3DCNN_encoder}
\end{figure}

The structure of the 3D-CNN to extract semantic features of cubic volumes in volume sequences is inspired by \cite{zreik2018recurrent}, as shown in Fig. \ref{fig_3DCNN_encoder}. The 3D-CNN consists of four sequentially connected substructures. Each substructure includes a convolutional layer with a convolution kernel size of $3\times3\times3$, a rectified linear unit (ReLU) and a $2\times2\times2$ max-pooling layer. The number of filters of the convolutional layer is 16 in the first part. In the remaining part, the number of filters of the convolutional layer is the number of the previous part multiplied by 2.

The feature maps obtained by the 3D-CNN are defined as $x\in \mathbb{R}^{C\times H\times H\times H}$, where $C$ and $H$ respectively indicate the number of filters and the size of feature maps. Since Transformer originates from NLP and needs to take a 1D vectors sequence as input, we flatten the feature maps into 1D vectors and arrange them into a sequence as the feature embeddings of Transformer.

\subsection{Transformer structure for global sequence analysis}
According to clinical experience, a coronary artery branch may have multiple plaques, and each plaque affects the blood flow velocity in the patient's coronary lumen. Therefore, analyzing the potential relationship between plaques at different locations is valuable for clinical diagnosis. In this work, for a voxel on the centerline, there is image information in both two directions (the direction of ascending aorta and the direction of the coronary end) that can affect the detection result.  To treat the feature maps of each coronary artery segment as the basis for judgment when detecting local coronary artery stenosis, we design Transformer structure to analyze feature sequence bidirectionally.

To introduce the order information of each cubic volume into our model, we add the learnable order embeddings\cite{dosovitskiy2020image} of the same dimension to feature embeddings before inputting embeddings into Transformer structure. The input embedding for Transformer structure $Z_{0}$ can be obtained by adding feature embeddings and order embeddings, expressed as follows:
\begin{equation}\label{eqn1}
	Z_{0}=[x_{1}+o_{1},x_{2}+o_{2},…,x_{L} +o_{L}]\in \mathbb{R}^{L\times ( C\cdot H^3)}
\end{equation}
where $x_{i}$ and $o_{i}$ respectively indicate the feature embedding and order embedding for the $i^{th}$ cubic volume.

The Transformer structure of TR-Net contains $T$ Transformer encoders, and the number $T$ of Transformer encoders is 12 in this work. Each Transformer encoder consists of two sub-blocks connected in sequence, multiheaded self-attention (MSA) and feed-forward network (FFN), where FFN consists of two linear layers with a ReLU activation. Layer normal (LN) and residual connections are respectively employed before and after both two sub-blocks\cite{baevski2018adaptive,wang2019learning}, as shown in Fig. \ref{fig_3DCNN_encoder}. For each Transformer encoder, the size of the input is the same as the output to ensure the consistency of Transformer encoders. The output of the previous Transformer encoder is treated as the input of the next Transformer encoder, the output of the $t^{th}$ Transformer encoder $Z_{t}$ can be defined as:
\begin{equation} \label{eqn2}
	\begin{split}
		Z_{t}’&= {\rm MSA}({\rm LN}(Z_{t-1}))\in \mathbb{R}^{L\times ( C\cdot H^3)}\\
		Z_{t} &= {\rm FFN}({\rm LN}(Z_{t}’) + Z_{t-1})) + Z_{t}’ +Z_{t-1}\in \mathbb{R}^{L\times ( C\cdot H^3)}
	\end{split}
\end{equation}
where $Z_{t-1}$ indicates the output of the ${t-1}^{th}$ Transformer encoder.
For the output of the last Transformer encoder $Z_{T}\in \mathbb{R}^{L\times(C\cdot H^3)}$, we split it into L embeddings, where the $i^{th}$ embedding is denoted as $Z_{T}^{i}\in \mathbb{R}^{1\times(C\cdot H^3)}$. The order of these embeddings corresponds to the order of the cubic volumes which input model. These embeddings are fed into softmax classifiers to detect significant stenosis of the corresponding cubic volume.

\section{Experiment}
\subsection{Dataset}
We conducted experiments on a dataset consisting of 76 CCTA scans from different patients and evaluated our method. These scan data have a total of 158 significant stenoses. We extracted the MPR image of the main coronary artery branches in each CCTA scan. For the entire dataset, we extracted a total of the MPR images of 609 coronary artery branches. For these MPR images, 42425 voxels belonging to the centerline of coronary artery branches could be selected as the volume center points. The dataset was annotated by experienced radiologists, and each voxel on the centerline was marked with non-significant stenosis (i.e. luminal narrowing $\le 50\%$) or significant stenosis (i.e. luminal narrowing $> 50\%$).

We selected voxels in the MPR image at intervals of 5 voxels along centerlines of coronary arteries and employed these voxels as volume center points to construct volume sequences. To extract local information properly, the side length $N$ of cubic volumes was set to 29 and the length $L$ of volume sequences was 30 at most. Considering that in most coronary artery branches, the proportion of significant stenosis in the entire branch is low, we cut the non-significant stenosis part of MPR images appropriately to make training samples as balanced as possible when constructing volume sequences. To ensure that the model has stronger robustness, we made the volume center points randomly move up to three voxels in any direction along 6 neighborhoods and rotated the cubic volumes at random angles perpendicular to the centerline. Finally, we obtained 849 volume sequences, of which 3326 center points corresponded to significant stenosis.

\begin{table}
	\Large
	\caption{Evaluation results for significant stenosis detection.}\label{tab1}
	\centering
	\setlength{\tabcolsep}{3.5mm}
	\resizebox{\textwidth}{!}{
		\begin{tabular}{l|l|l|l|l|l|l|l|l}
			\hline
			Method & Metric & ACC & Sens & Spec & PPV & NPV & F1 & MCC\\
			\hline
			Texture CLS\cite{tejero2019texture}& Orig data & 0.81 & 0.90 & 0.80 & \textbf{--} & \textbf{--} & \textbf{--} & \textbf{--} \\
			\hline
			3D-RCNN\cite{zreik2018recurrent} & Orig data & 0.94 & 0.63 & 0.97 & 0.65 & 0.97 & 0.64 & 0.60\\
			\hline
			2D-RCNN+PT\cite{denzinger2019coronary} & Orig data & 0.87 & 0.60 & 0.93 & 0.68 & 0.91 & 0.64 & 0.56\\
			\hline
			3D-RCNN\cite{zreik2018recurrent} & Our data & 0.87 & 0.66 & 0.91 & 0.56 & 0.93 & 0.60 & 0.53\\
			\hline
			2D-RCNN+PT\cite{denzinger2019coronary} & Our data & 0.89 & \textbf{0.82} & 0.89 & 0.50 & \textbf{0.97} & 0.62 & 0.58\\
			\hline
			TR-Net & Our data & \textbf{0.92} & 0.74 & \textbf{0.96} & \textbf{0.84} & 0.93 & \textbf{0.79} & \textbf{0.74}\\
			\hline
		\end{tabular}
	}
\end{table}

\subsection{Experimental Results}
Quantitative evaluation of experimental results can demonstrate the reliability of methods in clinical application, so we compared our proposed TR-Net with several state-of-the-art methods. To demonstrate the effectiveness of our TR-Net scientifically, we evaluated accuracy (ACC), sensitivity (Sens), specificity (Spec), predictive value (PPV), negative predictive value (NPV), F1-score, the Matthews correlation coefficient (MCC) based on the classification results. For all model experiments, we performed ten-fold cross-validation on centerline-level, where the validation set accounted for 10\% of the training data. Models were trained for 200 epochs, and the models were saved with the best performance on the validation set to predict the data on the test set. As shown in Table \ref {tab1}, compared with state-of-the-art methods, TR-Net achieved the best performance on ACC, Spec, PPV, F1 and MCC indicators on our dataset.

\begin{figure}
	\includegraphics[width=\textwidth]{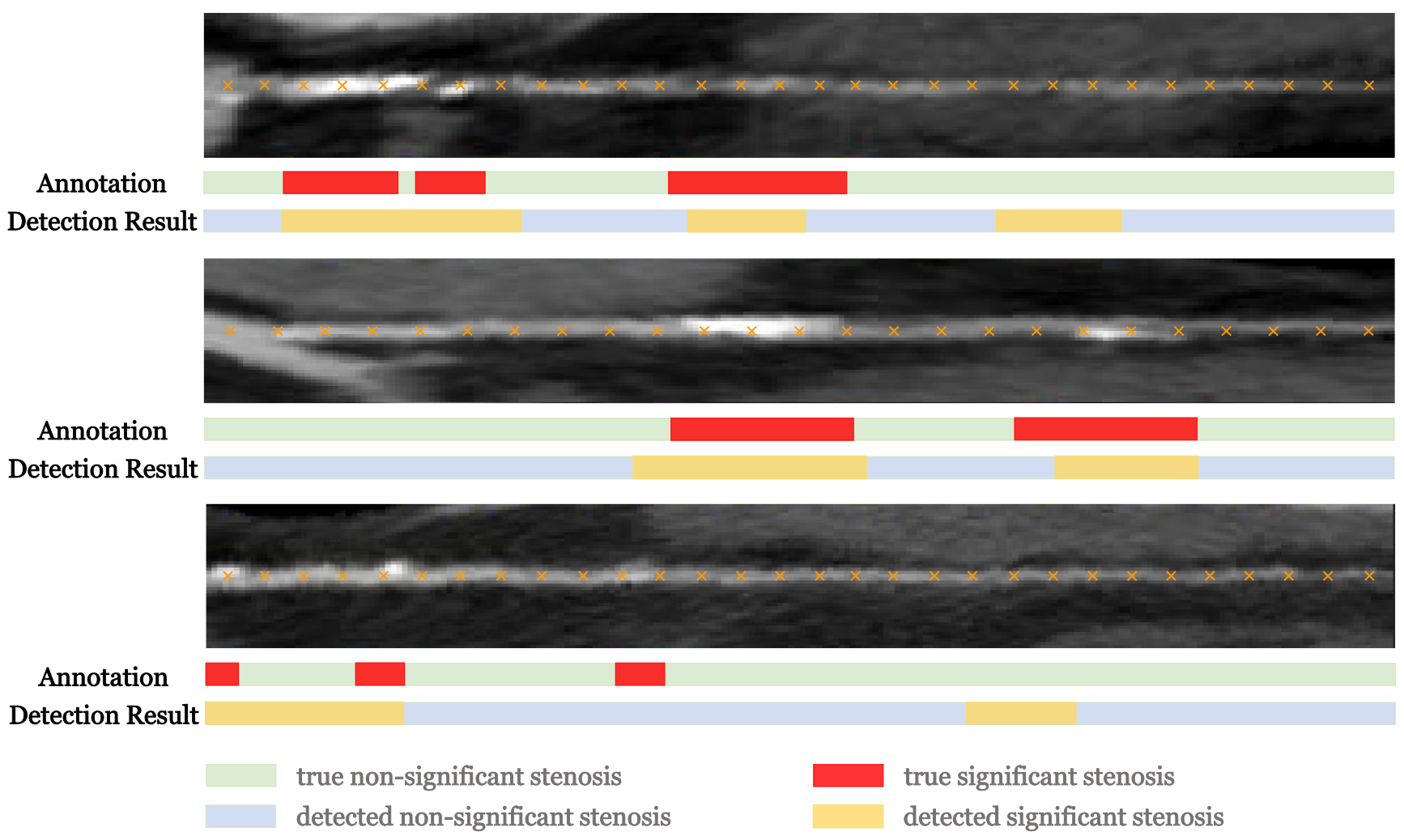}
	\caption{Examples of significant stenosis detection. Volume center points are denoted by orange $\times$. (Color figure online)} \label{fig_examples}
\end{figure}

The dataset we employed was obtained by marking whether each voxel was significant stenosis on the centerlines of coronary arteries. However, the model selected one for every 5 voxels along centerlines for significant stenosis detection. Therefore, there was a tolerable error between correct detection results and annotations at significant stenosis boundaries. If the error was less than 5 voxels, the detection results obtained by the model were considered correct. According to several representative examples of significant stenosis detection in Fig. \ref{fig_examples}, annotations and our detection results had high consistency. Experimental results demonstrated that TR-Net could effectively detect significant stenosis caused by various types of plaques, including non-calcified plaques that are difficult to detect. The detection results of our method had outstanding continuity, and there was almost no interruption when dealing with long-length significant stenosis.

\section{Conclusion}
In this work, we have proposed TR-Net to solve the challenging task of automatically detecting significant stenosis in MPR images. Our TR-Net can well combine the information of local areas adjacent to stenoses and the global information of coronary artery branches when detecting significant stenosis. Experimental results have demonstrated that TR-Net has better performance on multiple indicators compared with state-of-the-art methods. Through more comprehensive information analysis of coronary arteries in MPR images, our method can achieve the purpose of the computer-assisted diagnosis of CAD.

~\\
\textbf{Acknowledgments.} This work was supported by the National Natural Science Foundation of China under Grant 62001144 and Grant 62001141, and by China Postdoctoral Science Foundation under Grant 2021T140162 and Grant 2020M670911, and by Heilongjiang Postdoctoral Fund under Grant LBH-Z20066, and by Shandong Provincial Natural Science Foundation (ZR2020MF050).

%
%
%
%
\bibliographystyle{splncs}
\bibliography{mybibliography}
\end{document}